\documentclass[prb,aps,twocolumn,superscriptaddress,nofootinbib]{revtex4}

\renewcommand{\(}{\left(}
\renewcommand{\)}{\right)}
\newcommand{\vect}[1]{\mathbf{#1}}
\usepackage{amsmath}
\usepackage{pstricks}
\usepackage{pst-plot}
\usepackage{graphicx}
\usepackage{verbatim} 

\begin{document}
\title{Low-energy behavior of spin-liquid electron spectral functions}
\author{Evelyn Tang}
\affiliation{Department of Physics, Massachusetts Institute of Technology, Cambridge, MA 02139}

\author{Matthew P. A. Fisher}
\affiliation{Department of Physics, University of California, Santa Barbara, CA 93106}

\author{Patrick A. Lee}
\affiliation{Department of Physics, Massachusetts Institute of Technology, Cambridge, MA 02139}

\begin{abstract}

We calculate the electron spectral function for a spin-liquid with a spinon Fermi surface and a Dirac spin-liquid. Calculations are based upon the slave-rotor mean-field theory. We consider the effect of gauge fluctuations using a simple model and find the behavior is not strongly modified. The results, distinct from conventional Mott insulator or band theory predictions, suggest that measuring the spectral function e.g. via ARPES could help in the experimental verification and characterization of spin liquids.
\end{abstract}
\maketitle

\section{Introduction}
Anderson first proposed the resonating valence bond state in 1973 -- an example of a spin liquid: states with no preferred spin orientation and no broken symmetries. Such states may arise in frustrated spin systems such as the Heisenberg antiferromagnet on a kagome lattice. Only more recently have material candidates with promising experimental indications been found. The organic material\cite{organic} $\kappa\textrm{-(BEDT-TTF)}_2\textrm{Cu}_2\textrm{(CN)}_3$ and a related material based on $\textrm{EtMe}_3\textrm{Sb}[\textrm{Pd(dmit)}_2]_2$ molecules\cite{dmit} exhibit spin-liquid properties in the Mott insulating state, i.e. a lack of spin-order even down to very low temperatures. Further, their linear specific heat behaviour at low temperatures points toward the presence of a finite density of states of gapless excitations, whereas thermal transport measurements on the latter compound suggest these 
excitations are mobile\cite{dmit}.  Another compound Herbertsmithite $\textrm{ZnCu}_3\textrm{(OH)}_6\textrm{Cl}_2$ appears to have spin-liquid properties\cite{herbertst,herbertst2,herbertst3} with gapless spin excitations.

The organic materials are ``weak" Mott insulators which can be driven into a metallic phase under pressure.  
It has been proposed that the gapless excitations in such a weak Mott insulator can be modeled in terms of a Fermi sea of spinons - fermions that carry the spin, but not the charge, of the electron\cite{bosonmin,spinonsfs}.   Indeed, recent work has established strong evidence that such a spinon-Fermi sea state occurs in a triangular lattice Heisenberg model with four-site ring exchange, a model appropriate
for the organic Mott insulators\cite{Sheng1,Sheng2}.
An approach employing Fermionic spinons has also been proposed as an explanation for the
gapless behavior observed in the putative spin liquid phase in Herbertsmithite, except with the spinons filling a Dirac sea\cite{dslkagome,pifluxl,pifluxb}.

In addition to thermodynamic and transport experiments, signatures for such spin liquids might be expected 
in single electron properties such as those probed by angle-resolved photoemission spectroscopy (ARPES) or electron tunneling. 
In a conventional Mott insulator which exhibits antiferromagnetic order, the unit cell is doubled and the single hole spectrum can be described in terms of conventional band theory. For example, in high temperature superconductors, the holes occupy a band with a minimum at $(\pi/2,\pi/2)$\cite{arpescuprates,arpescuprateshf}. For a spin liquid the unit cell is not doubled, and band theory predicts a metal with a half-filled band, which completely fails to describes the Mott insulator. It follows that the single hole spectrum cannot have a conventional band-like description. Instead, within a slave-boson mean-field theory, the low-lying physical electron excitations are composites of a boson and the fermionic spinons and hence the spectral function behavior depends on the spinon dispersion. 

In this paper we work out the single electron spectral function for two models with different spinon spectra. Our goal is to stimulate experimental efforts to perform ARPES on spin-liquid candidate materials. The first model is one where the spinons form a Fermi surface (called a SFS spin liquid). There is considerable evidence that this state may describe the organic spin liquid materials.
A second model is the Dirac spin liquid (DSL) state. In particular  the $\pi$-flux phase on the kagome lattice has been proposed to describe Herbertsmithite\cite{dslkagome,pifluxl,pifluxb}. While more recent work supports a gapped spin liquid as the ground state of the $S=1/2$ antiferromagnetic Heisenberg model on the kagome lattice\cite{dmrg}, it is believed that many states very close in energy are competitive and the DSL remains an interesting state.

\section{Spectral function behavior\label{sec:spec}}
The Mott transition between a Fermi-liquid metal and a spinon Fermi sea spin-liquid insulator can be described with a slave-rotor formalism\cite{slaverotor}. Here the electron creation operator is written as a product of a charged spinless boson $b_i$ (``chargon") and a spin-1/2 charge-neutral ``spinon'' $f_{i\sigma}$, i.e. $c_{i\sigma}=f_{i\sigma}b_i$. 

\begin{figure}\begin{center}
\includegraphics[width=0.95\linewidth]{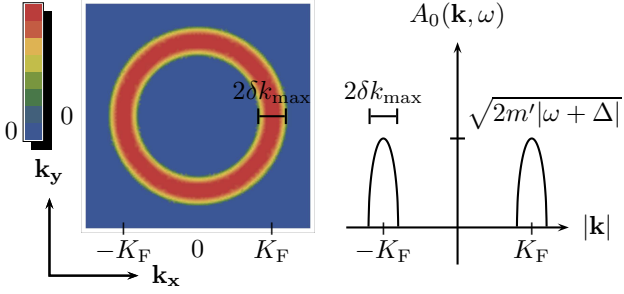}
\caption{(Color online). Spectral function for a SFS spin-liquid at constant energy (satisfying $|\omega|> \frac{\delta k^2}{2m_b} + \Delta $): excited electrons appear as a ring with radius $K_\textrm{F}$ in the Brillouin zone. The ring broadens as a function of energy --- ring width goes as $2\delta k_\textrm{max}=2\sqrt{2m_b|\omega+\Delta|}$. Maximum intensity is at $\vect{|k|}=K_\textrm{F}$ where $A_0(\vect{K_\textrm{F}},\omega)=\sqrt{2m'|\omega+\Delta|}$.}\label{fig:fsfixede2}
\end{center}
\end{figure}

At the mean-field level, in the metallic phase the bosons condense and the spinon Fermi surface becomes an electron Fermi surface with a quasiparticle weight proportional to $|\langle b_i\rangle|^2$. In the Mott insulator, the bosons are gapped while the spinons have a free hopping Hamiltonian. An emergent gauge field $a_\mu=(a_0,\vect{a})$ couples to both spinon and boson fields $f_\sigma$ and $b$, and the resulting theory\cite{mfspectral}  can be described by the action $S=\int_0^\beta d\tau\int d^2\vect{r} \mathcal{L}$, where
\begin{eqnarray}
\mathcal{L}&=&\mathcal{L}_b+\mathcal{L}_f+\mathcal{L}_g,\nonumber\\% +\mathcal{L}_{bf},\\
%\mathcal{L}_b&=&|(\partial_\mu+ia_\mu)b|^2+\Delta^2|b|^2,\nonumber\\
\mathcal{L}_b&=&|(\partial_\tau+ia_0)b|^2+|(v_b\vect{\nabla}+i\vect{a})b|^2+\Delta^2|b|^2,\nonumber\\
\mathcal{L}_f&=&f^\dag_\sigma(\partial_\tau-ia_0-\mu_f)f_\sigma+\frac{1}{2m_f}|(\vect{\nabla}-i\vect{a})f_\sigma|^2,\nonumber\\
\mathcal{L}_g&=&\frac{1}{4g^2}(\partial_\mu a_\nu-\partial_\nu a_\mu)^2  .
\end{eqnarray}
Here $v_b$ is the low-energy boson velocity, $\Delta$ the insulating gap and $m_f$ the effective spinon mass. Beyond mean-field the gauge-field fluctuations have to be considered, which will be discussed at the end of this paper.

At the mean-field level, the electron correlator factorizes locally (in space and time) into boson and spinon correlators. The retarded Green's function for the electron is then given by a convolution of boson and spinon Green's functions. At zero temperature we get\cite{mfspectral}
\begin{eqnarray}
G^R_0(\vect{k},\omega)
&=&\sum_{\vect{q},\alpha}\frac{M_{\alpha}(\vect{k-q},\vect{q})}{2\Omega_\vect{q}}\bigg[\frac{1-f(\xi^\alpha_{\vect{k-q}})}{\omega-\xi^\alpha_{\vect{k-q}}-\Omega_\vect{q}+i\eta}\nonumber\\
&&+ \frac{f(\xi^\alpha_{\vect{k-q}})}{\omega-\xi^\alpha_{\vect{k-q}}+\Omega_\vect{q}+i\eta}\bigg]  ,\label{eq:green0}
\end{eqnarray}
where $\xi^\alpha_{\vect{k-q}}$ is the dispersion of the $\alpha$ spinon band, $\Omega_\vect{q}$ is the boson dispersion and $f$ is the Fermi distribution function.

The form factor $M_{\alpha}(\vect{k-q},\vect{q})$ depends on the spinon and boson wavefunctions $\psi^\alpha$ and $\phi$,
\begin{eqnarray}
 M_{\alpha}(\vect{k-q},\vect{q})=\sum_{m}\psi_{m}^\alpha(\vect{k-q})\psi^{\alpha *}_m(\vect{k-q})\phi_{m}(\vect{q})\phi^*_{m}(\vect{q}) , \nonumber%\label{eq:ffsimple}
\end{eqnarray}
where $m$ runs over each atom in the basis. The derivation and evaluation of this expression is given in Appendix \ref{ff} --- at low-energies this form factor is simply a constant and does not change the qualitative result.

We obtain the electron spectral function as usual using $A_0(\vect{k},\omega)=-\frac{1}{\pi}\textrm{Im}G^R_0(\vect{k},\omega)$. For electron excitations (negative $\omega$) we keep only the second term in Eq. \ref{eq:green0}. At low-energies, the universal properties of the spectral function are determined only by the spinon and boson dispersions. For simplicity the boson dispersion is assumed quadratic,
\begin{eqnarray}
\Omega_\vect{q}-\Delta&=&\frac{\delta q^2}{2m_b}  ,%+O(\delta q^3)   
\end{eqnarray}
where $\Delta$ is the insulating gap and $m_b=\Delta/v_b^2$. Here $\vect{\delta q}$ is the boson momentum $\vect{q}$ measured from the band minimum or minima, which are located at $\vect{Q_b}$.

We consider two cases for the spinon dispersion. In the first there is a spinon Fermi surface and the second is a Dirac dispersion with a vanishing density of states at the chemical potential. 
\begin{figure}\begin{center}
\includegraphics[width=0.65\linewidth]{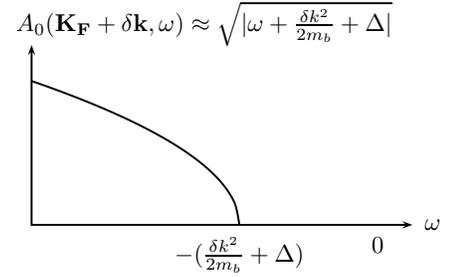}
\caption{Spectral function for a SFS spin-liquid at fixed momentum $\vect{k=K_F+\delta k}$ where $\vect{\delta k}$ is parallel to $\vect{K_F}$: we expect a square root dependence when varying $\omega$ (past the threshold energy of $|\omega_0|=\frac{\delta k^2}{2m_b}+\Delta$).}
\label{fig:fsfixedm}
\end{center}
\end{figure}

\subsection{Spin liquid with spinon Fermi surface (SFS)}
In the model proposed for the organic spin-liquid\cite{bosonmin,spinonsfs}, the spinons have a Fermi surface and the boson minimum is located at $\vect{Q_b}=0$. At low energies we obtain
\begin{eqnarray}
A_0(\vect{K_F+\delta k},\omega)
&=&\frac{1}{\Delta v_F}\sqrt{2m'|\omega+\frac{\delta k^2}{2m_b}+\Delta|};\nonumber\\ &&|\omega|> \frac{\delta k^2}{2m_b} + \Delta  , \label{eq:resultfs}
\end{eqnarray}
where $\vect{\delta k}$ is along $\vect{K_F}$ and $\frac{1}{m'}=\frac{1}{m_b}-\frac{1}{m_f}$. 
\begin{figure}\begin{center}
\includegraphics[width=0.95\linewidth]{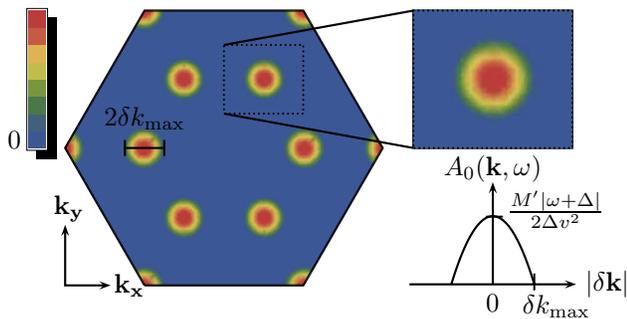}
\caption{(Color online). Spectral function for a DSL at constant energy (satisfying $|\omega|> \frac{\delta k^2}{2m_b} + \Delta $): excitations are expected at certain points $\{\vect{K_e}\}$ in the Brillouin zone. The dot radius size is a function of energy $\delta k_\textrm{max}=\sqrt{2m_b|\omega+\Delta|}$ and maximum intensity is at the center where $A_0(\vect{K_e},\omega)=\frac{M'}{2\Delta v^2}|\omega+\Delta|$. Note that here a kagome lattice with the $\pi$-flux state is shown for concreteness; however, a generic DSL will have the same behavior for low-energy excitations.}
\label{fig:diracbz}
\end{center}
\end{figure}
This has a square-root dependence on $|\omega|$ above the threshold excitation energy, $|\omega_0|=\frac{\delta k^2}{2m_b}+\Delta$. Spectral function measurements could detect such behavior, say using ARPES. When examining electronic excitations at constant energy (satisfying $|\omega|>\frac{\delta k^2}{2m_b}+\Delta$), one would measure a ring with radius $K_\textrm{F}$ in the Brillouin zone, see Fig. \ref{fig:fsfixede2}. The ring broadens as a function of energy, with ring width going as $2\delta k_\textrm{max}=2\sqrt{2m_b|\omega+\Delta|}$. The maximum intensity is at $\vect{|k|}=K_\textrm{F}$ where $A_0(\vect{K_\textrm{F}},\omega)=\sqrt{2m'|\omega+\Delta|}$.

If the experiment is done instead at fixed momentum $\vect{k}$, we expect a square root dependence when varying $\omega$ once electrons have enough energy to overcome the gap $\Delta$, see Fig. \ref{fig:fsfixedm}.

\subsection{Dirac spin liquid (DSL)}
For spinons with a Dirac spectrum, the dispersion is linear to lowest order in momentum
\begin{eqnarray}
\xi_{\vect{k}}&=&-v|\vect{\delta k}|  ,
\end{eqnarray}
where $v$ is the spinon velocity at the Dirac point and $\vect{\delta k}$ is $\vect{k}$ measured from the Dirac point.

After integration (details in Appendix \ref{sfintegral}), the spectral function becomes
\begin{eqnarray}
A_0(\vect{K_e+\delta k},\omega)
&=&\frac{M'}{2\Delta v^2}|\omega+\frac{\delta k^2}{2m_b}+\Delta|;\nonumber\\ &&|\omega|> \frac{\delta k^2}{2m_b} + \Delta  . \label{eq:result}
\end{eqnarray}
The constant $M'$ is the form factor contribution (defined and calculated in Appendices \ref{sfintegral} and \ref{ff}) which simplifies considerably near the Dirac point. 

The result has a linear dependence on $|\omega|$ past the threshold excitation energy, which we expect to see around a set of points of electron excitations $\vect{K_e}$ in the Brillouin zone. For concreteness we show this behavior for a $\pi$-flux state on the kagome lattice (details regarding this model are in the next section). 

\begin{figure}\begin{center}
\includegraphics[width=0.65\linewidth]{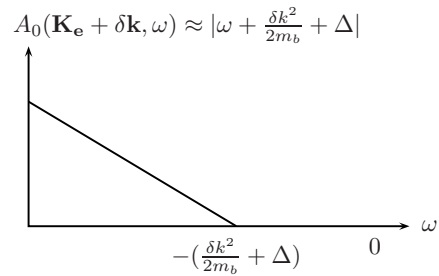}
\caption{Spectral function for a DSL at fixed momentum $\vect{k}$: we expect a linear dependence when varying $\omega$ past the threshold energy of $|\omega_0|=\frac{\delta k^2}{2m_b}+\Delta$ with slope $-\frac{M'}{2\Delta v^2}$.}
\label{fig:linear}
\end{center}
\end{figure}

In the hexagonal Brillouin zone of the kagome lattice, electron excitations are expected from eight points, see Fig. \ref{fig:diracbz}. Here we plot the spectral function at fixed energy. Once there is enough energy to overcome the gap $\Delta$, excitations appear at these points and broaden as a function of energy. The radius of these dots increases with energy $\delta k_\textrm{max}=\sqrt{2m_b|\omega+\Delta|}$ and maximum intensity would be observed at the center of these dots, $A_0(\vect{K_e},\omega)=\frac{M'}{2\Delta v^2}|\omega+\Delta|$. If a measurement is done at fixed momentum $\vect{k}$, we expect instead a linear dependence when varying $\omega$ with slope $-\frac{M'}{2\Delta v^2}$, see Fig. \ref{fig:linear}.

\section{$\pi$-flux state on the kagome lattice\label{sec:pi}}
As the results in the previous section are fairly general, we do not expect them to be strongly modified by microscopic details or the model used. The form-factor $M_{\alpha}(\vect{k-q},\vect{q})$ however, can be complicated and depends explicitly on band structure and lattice details. Here we pick a particular model to do an analytical expansion and calculation to make sure the form factor does not vanish or show additional singularities. 

Further, DSL models depend on a gauge choice that initially appears to break translation invariance. While spinon parameters do depend on the gauge chosen, the spectral function as a convolution of boson and spinon momenta is gauge-invariant (as expected for a physical observable).In our model we see explicitly that neither the gauge choice nor microscopic details affect our results. 

\begin{figure}\begin{center}
\includegraphics[width=\linewidth]{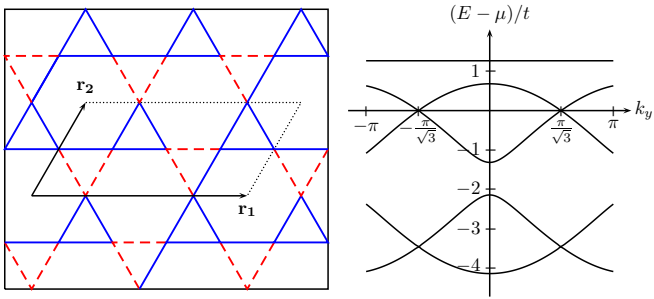}
\caption{(Color online). Kagome lattice with a doubled unit cell (translation vectors $\vect{r_1}$ and $\vect{r_2}$). Dashed lines correspond to bonds with negative hopping while unbroken lines correspond to bonds with positive hopping --- such that hexagons enclose a $\pi$-flux while triangles have zero flux. The band structure given by this hopping model (plotted along the line $k_x=0$) shows six bands (the top band is twofold degenerate). Spinon excitations are from the Dirac points at half-filling $\mu=(\sqrt{3}-1)t$ while boson excitations are from the lowest band minima.}\label{fig:kagome}
\end{center}
\end{figure}

The $\pi$-flux state on the kagome lattice was proposed as a candidate state for the Dirac spin liquid.\cite{pifluxl,pifluxb} In this model the tight-binding Hamiltonian describes a $\pi$-flux through every hexagon and zero flux through each triangle on the kagome lattice (see Fig. \ref{fig:kagome}). As this doubles the unit cell, the spinon and boson dispersions have six bands. 
With nearest-neighbor hopping $t$, we obtain the following dispersion (in units where the magnitude of the lattice spacing is set to 1/2)
\begin{eqnarray}
E_{\textrm{top}}=2t\qquad\textrm{(doubly degenerate)}\qquad\nonumber\\
E_{\pm,\mp}=t\(-1\pm\sqrt{3\mp\sqrt{2}\sqrt{3-\cos 2k_x+ 2\cos k_x\cos \sqrt{3}k_y}}\)\nonumber
\end{eqnarray}

At half-filling, the spinon chemical potential $\mu=(\sqrt{3}-1)t$ is located between the third $E_{+,-}$ and fourth $E_{+,+}$ bands which touch at two Dirac nodes $\vect{k}=(0,\pm\pi/\sqrt{3})$(Fig. \ref{fig:kagome}). For electron excitations, the spinon band $\alpha$ is just the fourth band. Bosons excitations are from the minima of the lowest band $E_{-,+}$ --- points $\vect{k}=\{\pm(\pi/3,0),\pm(2\pi/3,\pi/\sqrt{3})\}$ in the Brillouin zone. 

For low-energy excitations we expand to second order around a band minimum for bosons and a Dirac point for spinons,  
\begin{eqnarray}
&&\Omega_\vect{q}-\Delta=\frac{t_b}{4\sqrt{6}}\delta q^2+O(\delta q^3) , \\
&&\xi_{\vect{k}}=-\frac{t_f}{\sqrt{2}}|\vect{\delta k}|+ \frac{t_f}{4\sqrt{3}}(\vect{\delta k})^2+O(\delta k^3)   ,\nonumber
\end{eqnarray}
where $\vect{\delta q}$ and $\vect{\delta k}$ are momenta measured from the boson minimum and Dirac point respectively. Here the nearest-neighbor hopping $t$ has been replaced by $t_f$ and $t_b$ for effective spinon and boson hoppings obtained from the self-consistent mean-field solution. So in this model the boson mass and spinon velocity are $m_b=2\sqrt{6}/t_b$ and $v=t_f/\sqrt{2}$.

Since low energy excitations are composed of spinons near the Fermi energy and bosons near the band bottom, electron momenta will be the sum of both spinon and boson momenta. All possible combinations give rise to excitations centered around eight points in our halved Brillouin zone 
\begin{eqnarray}
\vect{K_e}=\(\pm\frac{\pi}{3},\pm\frac{\pi}{\sqrt{3}}\),\(\pm \frac{2\pi}{3},0\),\pm\(\frac{2\pi}{3},\frac{2\pi}{\sqrt{3}}\)     .   \label{eq:kpoints}
\end{eqnarray}
(The last two points can be translated to four other equivalent $\vect{K_e}$ points in the larger Brillouin zone using the original reciprocal lattice vectors.) 

Given the equivalent behavior of excitations in this system, it suffices to calculate the form-factor around one particular excitation where we find $M'\approx0.6$ (see Appendix \ref{ff}).

\begin{figure}\begin{center}
\includegraphics[width=\linewidth]{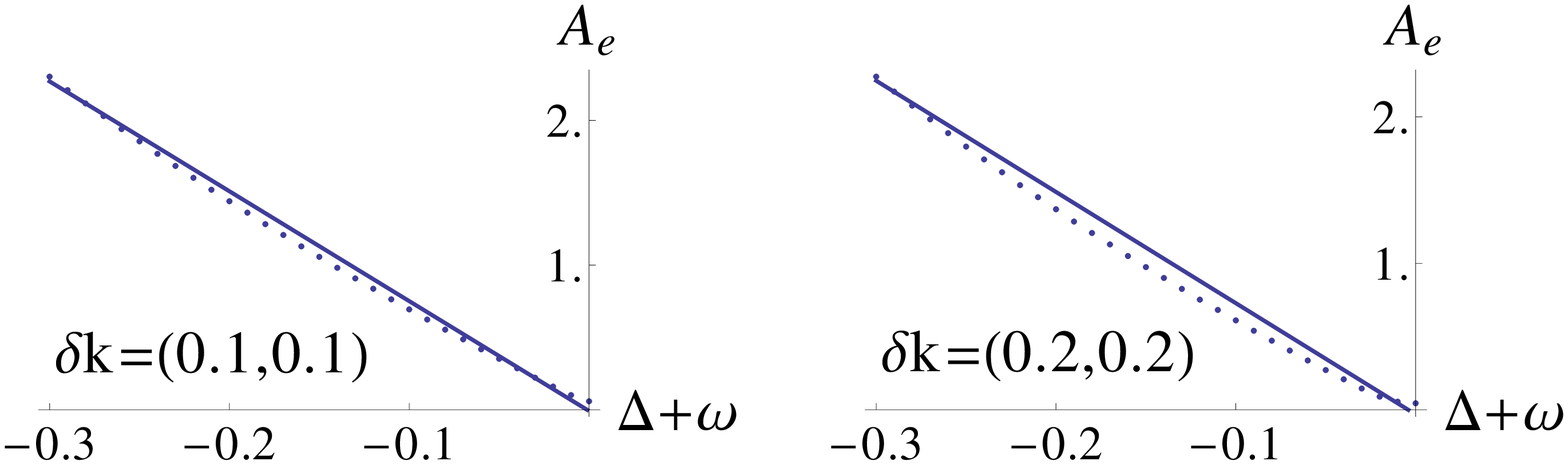}
\caption{(Color online). A comparison in our DSL $\pi$-flux model of the general linear prediction from Eq. \ref{eq:result} (straight line) with the actual mean-field behavior given by Eq. \ref{eq:spectral} (dotted lines). Here we choose low values of momenta close to one excitation point $\vect{k}\approx(\pi/3,\pi/\sqrt{3})$. With $t_b=t_f=1$, we compare $A_e(\vect{k},\omega)$ for two values of $\vect{\delta k}=(0.1,0.1)$ and $\vect{\delta k}=(0.2,0.2)$. We see the analytical expression is a good approximation to the mean-field expression at low energies, and is better for smaller values of $\vect{\delta k}$ as expected.}\label{fig:omega}
\end{center}
\end{figure}

We would like to compare the general linear prediction from Eq. \ref{eq:result} with the actual mean-field behavior of the spectral function in this DSL model.  
Here we choose low values of momenta close to one excitation point, $\vect{k}\approx(\pi/3,\pi/\sqrt{3})$. Setting $t_b=t_f=1$, we compare $A_e(\vect{k},\omega)$ for two values of $\vect{\delta k}=(0.1,0.1)$ and $\vect{\delta k}=(0.2,0.2)$ (see Fig. \ref{fig:omega}). We see the analytical expression is a good approximation to the mean-field expression at low energies, and is better for smaller values of $\vect{\delta k}$ as expected. 

\section{Effect of fluctuations}
Beyond mean-field, the dominant effect of gauge-field fluctuations is to bind the spinon and boson into an electron. Here we consider a simple model for these fluctuations to find the result does not change qualitatively for weak interaction strengths.

It is possible to approximate the attraction with an effective short-range interaction\cite{bindingl,bindingb} of the form
\begin{eqnarray}
- Uf^\dag_if_ib^\dag_ib_i, \qquad U>0  . \nonumber
\end{eqnarray}
This changes the electron Green's function to,
\begin{eqnarray}
G_e(\vect{k},\omega)=\frac{1}{G_0^{-1}(\vect{k},\omega)-U}\label{eq:totalGreen}   ,
\end{eqnarray}
where $G_0(\vect{k},\omega)$ is the mean-field expression given in Eq. \ref{eq:green0}. We denote its real and imaginary parts $G_0(\vect{k},\omega)=G'_0(\vect{k},\omega)+iG''_0(\vect{k},\omega)$. 

Having already calculated the imaginary part in Section \ref{sec:spec},
we need only to further calculate $G'_0(\vect{k},\omega)$.
In the following we focus on the Dirac spin-liquid.
As detailed in Appendix \ref{sec:bindingint}, in this case one finds,
\begin{eqnarray}
G'_0(\vect{k},\omega)
&=&\frac{1}{U_c}\(1-\frac{\tilde{\omega}}{q_c}\textrm{ln}\(\frac{q_c+\tilde{\omega}}{|\tilde{\omega}|}\)\) .\label{realGreen}
\end{eqnarray}
Here we have defined $\tilde{\omega}=(\omega+\frac{\delta k^2}{2m_b}+\Delta)/v$ and also $U_c=M'q_c/2\Delta $, with $q_c$  a high momentum cutoff. 

The electron spectral function in the presence of the attractive interaction between
the spinons and chargons can now be obtained by taking the imaginary part of $G_e(\vect{k},\omega)$ in Eq. \ref{eq:totalGreen},
\begin{eqnarray}
\textrm{Im}G_e(\vect{k},\omega)=\frac{G_0^{''}}{(1-UG_0')^2+(UG^{''}_0)^2}   .\nonumber
\end{eqnarray}
Notice that there is a pole only when $G_0^{''}=0$ (i.e. when $\tilde{\omega}>0$) and $G_0'=1/U$, so a transition occurs at the binding energy $U_c$.

When $U<U_c$, there is no bound state, so the presence of the spinon-chargon interaction merely alters the original linear behavior of $A_0(\vect{k},\omega)$.  To leading order
in $\tilde{\omega}$ the interacting electron spectral function becomes,
\begin{eqnarray}
A_e(\vect{k},\omega)&\approx&\frac{A_0(\vect{k},\omega)}{\(1-\frac{U}{U_c}\(1-\frac{\tilde{\omega}}{q_c}\textrm{ln}\(\frac{q_c+\tilde{\omega}}{\tilde{\omega}}\)\)\)^2}   ,\nonumber\\
&\approx&\frac{A_0(\vect{k},\omega)}{(1-U/U_c)^2}  ,\nonumber
\end{eqnarray}
i.e. $A_e(\vect{k},\omega)$ is still linear with an enhanced slope.

When $U>U_c$, there exists a bound state between the spinon and chargon (forming an electron) for $G_0'=1/U$, 
implying an electron state at energy
$\tilde{\omega}$ satisfying,
\begin{eqnarray}
\tilde{\omega}\textrm{ln}\(\frac{q_c+\tilde{\omega}}{\tilde{\omega}}\)=q_c\(1-\frac{U_c}{U}\)  .
\end{eqnarray}
The electron spectral function has a delta-function at this value of $\tilde{\omega}$ and a small incoherent part where $\tilde{\omega}<0$. The above equation can be solved numerically, e.g. for $U=2U_c$ and $q_c=1$ we get $\tilde{\omega}=0.4$. In this case $\omega=\tilde{\omega}v-\frac{\delta k^2}{2m_b}-\Delta$, i.e. the (negative) energy of the bound state within the gap decreases with $\delta k^2/2m_b$ and has its minimum on the Fermi surface. 

Hence, for weak interaction strength $U$, $A_e(\vect{k},\omega)$ has the same dependence on $\omega$ and $\vect{\delta k}$ as the mean-field result --- only the coefficient is modified.

\section{Discussion}
We present particular signatures in the electronic spectral function of spin-liquids that would distinguish such materials from a conventional Mott insulator or a conventional band theory description.  Using slave-rotor theory at the mean-field level, we consider two spin-liquid models: one with a spinon Fermi surface and another with a Dirac spinon sea, and extract the associated electron spectral functions.   We also explore the role of gauge fluctuations employing a simple model consisting of an attractive spinon-chargon interaction.  Solving this simple model indicates that
weak interactions do not change the qualitative behavior of the mean-field spectral functions,
but above a threshold interaction strength a spinon-chargon bound state (an electron) forms.
The energy of the bound state peels off from the continuum energy threshold, leading to a delta-function in the electron spectral function inside the gap.   We hope this work will encourage experimental efforts to perform ARPES on spin-liquid candidate materials. 

After completion of this paper, we learned that Pujari and Lawler also studied the ARPES signature of the DSL state on the kagome lattice, to find the linear onset is unaffected by gauge fluctuations to first order\cite{lawler}.
%At present we do not know the effect of gauge fluctuations. If they are sufficiently weak, the mean-field behavior we predict is not modified. However, it is certainly possible gauge fluctuations are sufficiently strong that qualitatively different behavior arises. 

We thank Andrew Potter and David Mross for helpful discussions.  This work was supported by the NSF under grants DMR-1104498 (P.A.L.) and DMR-1101912 (M.P.A.F.) and by the Caltech Institute
of Quantum Information and Matter, an NSF Physics
Frontiers Center with support of the Gordon and Betty Moore
Foundation (M.P.A.F.).

\appendix
\section{DSL spectral function integral\label{sfintegral}}
Here we evaluate the spectral function
\begin{eqnarray}
A_0(\vect{k},\omega)
&=&\sum_{\vect{q},\alpha}\frac{1}{2\Omega_\vect{q}}(M_{\alpha}(\vect{k-q},\vect{q})\delta(\omega-\xi^\alpha_{\vect{k-q}}-\Omega_\vect{q})\nonumber\\
&&+M_{\alpha}(\vect{k-q},\vect{q})\delta(\omega-\xi^\alpha_{\vect{k-q}}+\Omega_\vect{q}))   ,\label{eq:spectral}
\end{eqnarray}
for a Dirac spin liquid.

As the boson wavefunctions are quadratic at the band minima, $M(\vect{k-q},\vect{q})$ depends only on spinon momenta. Taking $\alpha$ to be the spinon band below the Fermi energy, 
\begin{eqnarray}
&&A_0(\vect{k},\omega)\nonumber\\&\approx&\sum_{\vect{q}}\frac{1}{2\Delta}M(\vect{\delta k-\delta q})\delta(\omega+v|\vect{\delta k-\delta q}|\nonumber\\&&-\frac{v}{2\sqrt{6}}(\delta k-\delta q)^2+\frac{\delta q^2}{2m_b}+\Delta)\nonumber\\
&=&\sum_{\vect{q'}}\frac{M(\vect{q'})}{2\Delta}\delta(\omega+v|\vect{q'}|-\frac{vq'^2}{2\sqrt{6}}+\frac{(\delta k-q')^2}{2m_b}+\Delta)   , \nonumber
\end{eqnarray}
where we changed coordinates to $\vect{q'}=\vect{\delta k}-\vect{\delta q}$, momenta about the Dirac points.

As the low-energy spinon wavefunctions do not depend linearly on $|q'|$ but only on $\theta$ --- the direction of $\vect{q'}$ around the Dirac point --- $M(q')$ reduces to $M(\theta)$ (independent of the magnitude of $\vect{\delta k}-\vect{\delta q}$). In polar coordinates the integral becomes
\begin{eqnarray}
A_0(\vect{k},\omega)
&=&\frac{1}{2\Delta}\int^{2\pi}_0 d\theta M(\theta)\int_0^{q_c} dq' q' \delta(\omega+v|q'|\nonumber\\
&&-\frac{vq'^2}{2\sqrt{6}}+\frac{(\delta k-q')^2}{2m_b}+\Delta)\nonumber\\
&\approx&\frac{1}{2\Delta}\int^{2\pi}_0 d\theta M(\theta)\int_0^{q_c} dq' q' \delta(vq'+\omega+\frac{\delta k^2}{2m_b}+\Delta)\nonumber\\
&=&\frac{M'}{2\Delta v^2}|\omega+\frac{\delta k^2}{2m_b}+\Delta|;\qquad |\omega+\frac{\delta k^2}{2m_b}| > \Delta   , \label{eq:final}
\end{eqnarray}
where we define the constant $M'=\int^{2\pi}_0 d\theta M(\theta)$.

\section{Form-factor contribution $M'$\label{ff}}
$M_{\alpha}(\vect{k-q},\vect{q})$, the ``form factor'' given in Section \ref{sec:spec} depends on the spinon and boson wavefunctions $\psi^\alpha$ and $\phi$ (which have $m$ indices for each atom in the basis) --- so is an $m\times m$ matrix,
\begin{eqnarray}
&&M_{\alpha,mn}(\vect{k-q},\vect{q})=\nonumber\\
&&\sum_{l=0,3}e^{i\vect{k}\cdot (\vect{r_m}-\vect{r_n})}\psi_{m+l}^\alpha(\vect{k-q})\psi^{\alpha *}_{n}(\vect{k-q})\phi_{m+l}(\vect{q})\phi^*_{n}(\vect{q})  . \nonumber%\label{eq:ff}
\end{eqnarray}
Since the $\pi$-flux state has a doubled unit cell, we sum over $l=0,3$ to double the number of atoms as $m,n=1,2,3$ in the original basis.

Resolution on an atomic level is unecessary so we take the trace,
\begin{eqnarray}
M_{\alpha}(\vect{k-q},\vect{q})&=& \textrm{Tr}M_{\alpha,mn}(\vect{k-q},\vect{q})\nonumber\\
&=&\sum_{m}M_{\alpha,mm}(\vect{k-q},\vect{q})   .   \nonumber
\end{eqnarray}
To simplify the expression further, we use certain properties of the model that also demonstrate the explicit translation invariance (in the original unit cell) of our model and hence its gauge invariance. 

Each electron excitation has contributions from the two spinon Dirac points. The Hamiltonian at the first Dirac point is the complex conjugate of the Hamiltonian at the second Dirac point,
\begin{eqnarray}
H(\vect{k}=(0,\pi/\sqrt{3}))=H^*(\vect{k}=(0,-\pi/\sqrt{3}))  , \nonumber 
\end{eqnarray}

Also, at low energies a spinon from the first Dirac point combines only with bosons from the two minima $\vect{q}\approx\{(\pi/3,0),(-2\pi/3,-\pi/\sqrt{3})\}$, while spinons from the second Dirac point combine only with bosons from a non-overlapping set of minima, $\vect{q}\approx\{(-\pi/3,0),(2\pi/3,\pi/\sqrt{3})\}$. The boson wavefunctions that pair with a spinon from the first Dirac point is the complex conjugate of boson wavefunctions that pair with the second:%spinon from the second Dirac point: 
\begin{eqnarray}
(\phi_{m+l}(\vect{q})\phi^*_{n}(\vect{q}))^*=\phi_{n+l}(\vect{q}\pm(\pi,\pi/\sqrt{3}))\phi^*_{m}(\vect{q}\pm(\pi,\pi/\sqrt{3}))\nonumber\label{eq:bosoncc}
\end{eqnarray}

These properties enable us to simplify the form factor --- as each excitation has two contributions that are complex conjugates, their imaginary part cancels --- 
\begin{eqnarray}
&&M_{\alpha}(\vect{k-q},\vect{q})\nonumber\\
&=&\sum_{m,l}\psi_{m+l}^\alpha(\vect{k-q})\psi^{\alpha *}_m(\vect{k-q})\phi_{m+l}(\vect{q})\phi^*_{m}(\vect{q})\nonumber\\
&=&\sum_{m}|\psi_{m}^\alpha(\vect{k-q})|^2|\phi_{m}(\vect{q})|^2\label{eq:realm}\\&&+\textrm{Re}[\psi_{m+l}^\alpha(\vect{k-q})\psi^{\alpha *}_{m}(\vect{k-q})\phi_{m+l}(\vect{q})\phi^*_{m}(\vect{q})]\nonumber
\end{eqnarray}
\textit{As the form factor is now explicitly Hermitian, it preserves translation invariance in the original unit cell and also gauge-invariance.}

\begin{figure}\begin{center}
\includegraphics[width=0.8\linewidth]{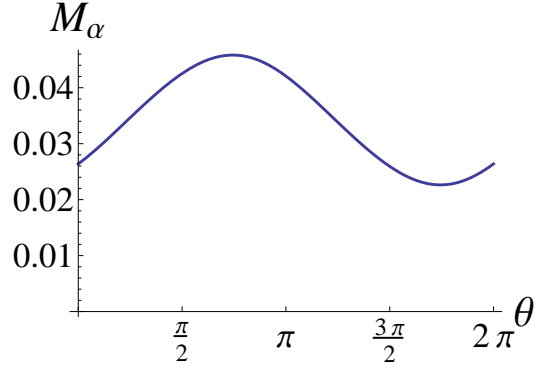}
\caption{A plot of $M_{\alpha}(\theta)$ against $\theta$, it is a smoothly varying function.}\label{fig:rangeall}
\end{center}
\end{figure}

We now calculate $M'$ for an excitation near $\vect{K_e}=(\pi/3,\pi/\sqrt{3})$ with a spinon from $\vect{k-q}\approx(0,\pi/\sqrt{3})$. At low-energies, $\alpha$ is the fourth band and we expand the Hamiltonian to a subspace of the two bands near half-filling. % using first order perturbation theory. %using eigenvectors from our tight-binding Hamiltonian, 
Computing $M(\theta)$ (see Fig. \ref{fig:rangeall}), this is
\begin{eqnarray}
&&M'\nonumber\\
&=&\int^{2\pi}_0 d\theta M(\theta)\nonumber\\
&=&\int^{2\pi}_0 d\theta \sum_{m}M_{mm}(\theta)\nonumber\\
&=&\int^{2\pi}_0 d\theta\sum_{m=1,2,3}\{|\psi_{m}(\theta)|^2|\phi_{m}|^2\nonumber\\&+&\textrm{Re}[\psi_{3+m}(\theta)\psi^*_{m}(\theta)\phi_{3+m}\phi^*_{m}]\}\nonumber\\
&=&\int_0^{2\pi} d\theta\{|\psi_{1}(\theta)|^2|\phi_{1}|^2+|\psi_{2}(\theta)|^2|\phi_{2}|^2+|\psi_{3}(\theta)|^2|\phi_{3}|^2\nonumber\\
&+&\textrm{Re}[\psi_{4}(\theta)\psi^*_{1}(\theta)\phi_{4}\phi^*_{1}+\psi_{5}(\theta)\psi^*_{2}(\theta)\phi_{5}\phi^*_{2}%\nonumber\\&&
+\psi_{6}(\theta)\psi^*_{3}(\theta)\phi_{6}\phi^*_{3}]\}\nonumber\\
&=& 0.604
\end{eqnarray}
so the form factor $M'$ suppresses the spectral function by about $2\pi/0.604\approx10$.

Hence, the form factor merely enhances or suppresses particular excitations as a smoothly varying function of angle around the Dirac point and is independent of $\omega$ or the magnitude of $\vect{k}$ in this expansion.

\section{Binding energy integral\label{sec:bindingint}}
We can use a Kramers-Kronig relation to obtain the real part of our mean-field Green's function from Eq. \ref{eq:green0}
\begin{align}
&G'_0(\vect{k},\omega)\nonumber=\\
&\sum_{\vect{q},\alpha}\frac{1}{2\Omega_\vect{q}}\mathcal{P}\bigg(\frac{M_{\alpha}(\vect{k-q},\vect{q})}{\omega-\xi^\alpha_{\vect{k-q}}-\Omega_\vect{q}}
+\frac{M_{\alpha}(\vect{k-q},\vect{q})}{\omega-\xi^\alpha_{\vect{k-q}}+\Omega_\vect{q}}\bigg)\label{eq:realgex}
\end{align}
Keeping only electron excitations and expanding at low energies, we get
\begin{align}
&G'_0(\vect{k},\omega\approx\nonumber\\
&\sum_{\vect{q}}\frac{1}{2\Delta}\mathcal{P}\bigg(\frac{M(\vect{\delta k-\delta q})}{\omega+v|\vect{\delta k-\delta q}|-\frac{v(\delta k-\delta q)^2}{2\sqrt{6}}+\frac{\delta q^2}{2m_b}+\Delta}\bigg)
\end{align}
Changing again to $\vect{q'}=\vect{\delta k}-\vect{\delta q}$ and using polar coordinates this becomes
\begin{eqnarray}
&&G'_0(\vect{k},\omega)\nonumber\\
&=&\frac{1}{2\Delta}\int d\theta M(\theta)\int dq' q' \mathcal{P}\(\frac{1}{\omega+v|q'|-\frac{vq'^2}{2\sqrt{6}}+\frac{(\delta k-q')^2}{2m_b}+\Delta}\)\nonumber\\
&\approx&\frac{1}{2\Delta}\int d\theta M(\theta)\int dq' q'\mathcal{P}\(\frac{1}{vq'+\omega+\frac{\delta k^2}{2m_b}+\Delta}\)\nonumber\\
&=&\frac{M'}{2\Delta v}\int^{q_c}_0 dq q \mathcal{P}\(\frac{1}{q+(\omega+\frac{\delta k^2}{2m_b}+\Delta)/v}\)
\end{eqnarray}
where $q_c$ is a cutoff momenta far above the scale of low energy excitations.

To evaluate this integral, we look at $\tilde{\omega}=(\omega+\frac{\delta k^2}{2m_b}+\Delta)/v>0$ and $\tilde{\omega}<0$.

For $\tilde{\omega}>0$:
\begin{eqnarray}
\int^{q_c}_0 dq q \mathcal{P}\(\frac{1}{q+\tilde{\omega}}\)
&=& \int^{q_c}_0 dq \(1-\frac{\tilde{\omega}}{q+\tilde{\omega}}\)\nonumber\\
&=& q_c-\tilde{\omega}\textrm{ln}\(\frac{q_c+\tilde{\omega}}{\tilde{\omega}}\)%\label{negom}
\end{eqnarray}

For $\tilde{\omega}<0$:
\begin{eqnarray}
&&\int^{q_c}_0 dq q \mathcal{P}\(\frac{1}{q+\tilde{\omega}}\)\nonumber\\
&=& \int^{|\tilde{\omega}|-\eta}_0 \frac{dq q}{q+\tilde{\omega}}+ \int^{q_c}_{|\tilde{\omega}|+\eta} \frac{dq q}{q+\tilde{\omega}}\nonumber\\
&=& |\tilde{\omega}|-\eta-\tilde{\omega}\textrm{ln}\(\frac{\eta}{|\tilde{\omega}|}\)+q_c-(|\tilde{\omega}|+\eta)-\tilde{\omega}\textrm{ln}\(\frac{q_c+\tilde{\omega}}{\eta}\)\nonumber\\
&=& q_c-\tilde{\omega}\textrm{ln}\(\frac{q_c+\tilde{\omega}}{|\tilde{\omega}|}\)%\label{posom}
\end{eqnarray}
Combining both expressions, we obtain the real part of the Green's function
\begin{eqnarray}
G'_0(k,\omega)
&=&\frac{M'}{2\Delta v}\(q_c-\tilde{\omega}\textrm{ln}\(\frac{q_c+\tilde{\omega}}{|\tilde{\omega}|}\)\)\nonumber\\
&=&\frac{1}{U_c}\(1-\frac{\tilde{\omega}}{q_c}\textrm{ln}\(\frac{q_c+\tilde{\omega}}{|\tilde{\omega}|}\)\)
\end{eqnarray}
where we define $U_c=M'q_c/2\Delta $.

\end{document}